\input epsf
\newfam\scrfam
\batchmode\font\tenscr=rsfs10 \errorstopmode
\ifx\tenscr\nullfont
        \message{rsfs script font not available. Replacing with calligraphic.}
        \def\scr{\cal}
\else   
        \font\sevenscr=rsfs7
        \font\fivescr=rsfs5
        \skewchar\tenscr='177 \skewchar\sevenscr='177 \skewchar\fivescr='177
        \textfont\scrfam=\tenscr \scriptfont\scrfam=\sevenscr
        \scriptscriptfont\scrfam=\fivescr
        \def\scr{\fam\scrfam}
        \def\cal{\scr}
\fi
\catcode`\@=11
\newfam\frakfam
\batchmode\font\tenfrak=eufm10 \errorstopmode
\ifx\tenfrak\nullfont
        \message{eufm font not available. Replacing with italic.}
        
\else
	
	\font\sevenfrak=eufm7 \font\fivefrak=eufm5
	\textfont\frakfam=\tenfrak
	\scriptfont\frakfam=\sevenfrak \scriptscriptfont\frakfam=\fivefrak
	
\fi
\catcode`\@=\active
\newfam\msbfam
\batchmode\font\twelvemsb=msbm10 scaled\magstep1 \errorstopmode
\ifx\twelvemsb\nullfont\def\Bbb{\bf}

	\message{Blackboard bold not available. Replacing with boldface.}
\else   \catcode`\@=11
        \font\tenmsb=msbm10 \font\sevenmsb=msbm7 \font\fivemsb=msbm5
        \textfont\msbfam=\tenmsb
        \scriptfont\msbfam=\sevenmsb \scriptscriptfont\msbfam=\fivemsb
        \def\Bbb{\relax\expandafter\Bbb@}
        \def\Bbb@#1{{\Bbb@@{#1}}}
        \def\Bbb@@#1{\fam\msbfam\relax#1}
        \catcode`\@=\active

\fi
        \font\eightrm=cmr8              \def\xrm{\eightrm}
        \font\eightbf=cmbx8             \def\xbf{\eightbf}
        \font\eightit=cmti10 at 8pt     \def\xit{\eightit}
        \font\eighttt=cmtt8             \def\xtt{\eighttt}
        \font\eightcp=cmcsc8
        \font\eighti=cmmi8              \def\xold{\eighti}
        \font\eightib=cmmib8             \def\xbold{\eightib}
        \font\teni=cmmi10               \def\old{\teni}
        \font\tencp=cmcsc10
        \font\tentt=cmtt10
        
        \font\twelvecp=cmcsc10 scaled\magstep1

	 at10pt	
	\font\twelvehelvbold=phvb at12pt
	 at14pt
	\font\sixteenhelvbold=phvb at16pt

\def\noblackbox{\overfullrule=0pt}
\noblackbox

\newtoks\headtext
\headline={\ifnum\pageno=1\hfill\else
	\ifodd\pageno{\eightcp\the\headtext}{ }\dotfill{ }{\old\folio}
	\else{\old\folio}{ }\dotfill{ }{\eightcp\the\headtext}\fi
	\fi}
\def\makeheadline{\vbox to 0pt{\vss\noindent\the\headline\break
\hbox to\hsize{\hfill}}
        \vskip2\baselineskip}
\newcount\infootnote
\infootnote=0
\def\foot#1#2{\infootnote=1
\footnote{${}^{#1}$}{\vtop{\baselineskip=.75\baselineskip
\advance\hsize by -\parindent\noindent{\xrm #2}}}\infootnote=0$\,$}
\newcount\refcount
\refcount=1
\newwrite\refwrite
\def\oldsize{\ifnum\infootnote=1\xold\else\old\fi}
\def\ref#1#2{
	\def#1{{{\oldsize\the\refcount}}\ifnum\the\refcount=1\immediate\openout\refwrite=\jobname.refs\fi\immediate\write\refwrite{\item{[{\xold\the\refcount}]} 
	#2\hfill\par\vskip-2pt}\xdef#1{{\noexpand\oldsize\the\refcount}}\global\advance\refcount by 1}
	}
\def\refout{\catcode`\@=11
        \xrm\immediate\closeout\refwrite
        \vskip2\baselineskip
        {\noindent\twelvecp References}\hfill\vskip\baselineskip
        \baselineskip=.75\baselineskip
        \input\jobname.refs
        \baselineskip=4\baselineskip \divide\baselineskip by 3
        \catcode`\@=\active\rm}

\def\hepth#1{\href{http://xxx.lanl.gov/abs/hep-th/#1}{{\xtt hep-th/#1}}}
\def\jhep#1#2#3#4{\href{http://jhep.sissa.it/stdsearch?paper=#2\%28#3\%29#4}{J. High Energy Phys. {\xbold #1#2} ({\xold#3}) {\xold#4}}}
\def\AP#1#2#3{Ann. Phys. {\xbold#1} ({\xold#2}) {\xold#3}}

\def\CQG#1#2#3{Class. Quantum Grav. {\xbold#1} ({\xold#2}) {\xold#3}}

\def\JHEP{\jhep}

\def\NPB#1#2#3{Nucl. Phys. {\xbf B}{\xbold#1} ({\xold#2}) {\xold#3}}

\def\PLB#1#2#3{Phys. Lett. {\xbf B}{\xbold#1} ({\xold#2}) {\xold#3}}

\newcount\sectioncount
\sectioncount=0
\def\section#1#2{\global\eqcount=0
	\global\subsectioncount=0
        \global\advance\sectioncount by 1
	\ifnum\sectioncount>1
	        \vskip2\baselineskip
	\fi
	\noindent
        \line{\twelvecp\the\sectioncount. #2\hfill}
		\vskip.8\baselineskip\noindent
        \xdef#1{{\old\the\sectioncount}}}
\newcount\subsectioncount
\def\subsection#1#2{\global\advance\subsectioncount by 1
	\vskip.8\baselineskip\noindent
	\line{\tencp\the\sectioncount.\the\subsectioncount. #2\hfill}
	\vskip.5\baselineskip\noindent
	\xdef#1{{\old\the\sectioncount}.{\old\the\subsectioncount}}}
\newcount\appendixcount
\appendixcount=0
\def\appendix#1{\global\eqcount=0
        \global\advance\appendixcount by 1
        \vskip2\baselineskip\noindent
        \ifnum\the\appendixcount=1
        \hbox{\twelvecp Appendix A: #1\hfill}\vskip\baselineskip\noindent\fi
    \ifnum\the\appendixcount=2
        \hbox{\twelvecp Appendix B: #1\hfill}\vskip\baselineskip\noindent\fi
    \ifnum\the\appendixcount=3
        \hbox{\twelvecp Appendix C: #1\hfill}\vskip\baselineskip\noindent\fi}
\def\acknowledgements{\vskip2\baselineskip\noindent
        \underbar{\it Acknowledgements:}\ }
\newcount\eqcount
\eqcount=0
\def\Eqn#1{\global\advance\eqcount by 1
\ifnum\the\sectioncount=0
	\xdef#1{{\old\the\eqcount}}
	\eqno({\oldstyle\the\eqcount})
\else
        \xdef#1{{\old\the\sectioncount}.{\old\the\eqcount}}
        \ifnum\the\appendixcount=0
                \eqno({\oldstyle\the\sectioncount}.{\oldstyle\the\eqcount})\fi
        \ifnum\the\appendixcount=1
                \eqno({\oldstyle A}.{\oldstyle\the\eqcount})\fi
        \ifnum\the\appendixcount=2
                \eqno({\oldstyle B}.{\oldstyle\the\eqcount})\fi
        \ifnum\the\appendixcount=3
                \eqno({\oldstyle C}.{\oldstyle\the\eqcount})\fi
\fi}
\def\eqn{\global\advance\eqcount by 1
\ifnum\the\sectioncount=0
	\eqno({\oldstyle\the\eqcount})
\else
        \ifnum\the\appendixcount=0
                \eqno({\oldstyle\the\sectioncount}.{\oldstyle\the\eqcount})\fi
        \ifnum\the\appendixcount=1
                \eqno({\oldstyle A}.{\oldstyle\the\eqcount})\fi
        \ifnum\the\appendixcount=2
                \eqno({\oldstyle B}.{\oldstyle\the\eqcount})\fi
        \ifnum\the\appendixcount=3
                \eqno({\oldstyle C}.{\oldstyle\the\eqcount})\fi
\fi}
\def\multi{\global\advance\eqcount by 1}
\def\multieq#1#2{\xdef#1{{\old\the\eqcount#2}}
        \eqno{({\oldstyle\the\eqcount#2})}}
\newtoks\url
\def\Href#1#2{\catcode`\#=12\url={#1}\catcode`\#=\active#2}
\def\href#1#2{{#2}}

\parskip=3.5pt plus .3pt minus .3pt
\baselineskip=14pt plus .1pt minus .05pt
\lineskip=.5pt plus .05pt minus .05pt
\lineskiplimit=.5pt
\abovedisplayskip=18pt plus 4pt minus 2pt
\belowdisplayskip=\abovedisplayskip
\hsize=14cm
\vsize=19cm
\hoffset=1.5cm
\voffset=1.8cm
\frenchspacing
\footline={}
\def\ts{\textstyle}
\def\ss{\scriptstyle}

\def\*{\partial}
\def\punkt{\,\,.}
\def\komma{\,\,,}

\def\={\!=\!}
\def\small#1{{\hbox{$#1$}}}
\def\half{\small{1\over2}}
\def\fraction#1{\small{1\over#1}}
\def\fr{\fraction}
\def\Fraction#1#2{\small{#1\over#2}}
\def\Fr{\Fraction}

\def\eg{{\tenit e.g.}}

\def\ie{{\tenit i.e.}}

\def\nlni{\hfill\break}

\def\a{\alpha}
\def\b{\beta}

\def\d{\delta}

\def\g{\gamma}
\def\l{\lambda}

\def\G{\Gamma}

\def\H{{\Bbb H}}

\def\ra{\rightarrow}

\def\Dslash{D\hskip-6.5pt/\hskip1.5pt}

%
%

\def\D{\Delta}

\def\G{\Gamma}

\def\a{\alpha}
\def\b{\beta}

\def\H{{\Bbb H}}

\def\N{{\cal N}}

\def\II{\hbox{I\hskip-0.6pt I}}

\def\lra{\longrightarrow}
\def\ra{\rightarrow}
\def\sea{\searrow}
\def\sa{\downarrow}
\def\swa{\swarrow}

\def\arrowunder#1{\raise4pt\vtop{\baselineskip=0pt\lineskip=0pt
      \ialign{\hfill##\hfill\cr${\ss #1}$\cr$\lra$\cr}}}

\def\Darrow#1{\;\arrowunder{\D_{#1}}\;}

\def\fr{\fraction}
\def\Fr{\Fraction}

\def\H{{\cal H}}

%
%
%

\headtext={M. Cederwall, B.E.W.Nilsson and D. Tsimpis: 
``Spinorial cohomology...''}

\null\vskip-2cm
\line{
\epsfysize=1.7cm
\epsffile{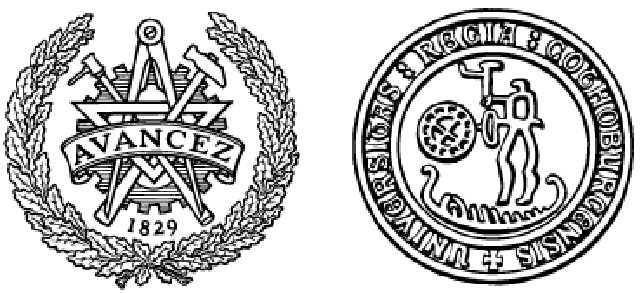}
\hfill}
\vskip-1.7cm
\line{\hfill G\"oteborg ITP preprint}
\line{\hfill\tt hep-th/0110069}
\line{\hfill October, {\old2001}}
\line{\hrulefill}

\vfill

\centerline{\sixteenhelvbold Spinorial cohomology} 
\vskip6pt 
\centerline{\sixteenhelvbold and maximally supersymmetric theories}

\vskip1.6cm

\centerline{\twelvehelvbold Martin Cederwall,
	 Bengt E.W. Nilsson and Dimitrios Tsimpis}

\vskip.8cm

\centerline{\it Department of Theoretical Physics}
\centerline{\it G\"oteborg University and Chalmers University of Technology }
\centerline{\it SE-412 96 G\"oteborg, Sweden}

\vskip1.6cm

\noindent\underbar{Abstract:} Fields in supersymmetric gauge theories
may be seen as elements in a spinorial cohomology. We elaborate on this
subject, specialising to maximally supersymmetric theories, where
the superspace Bianchi identities, after suitable conventional constraints
are imposed, put the theories on shell. In these cases, the spinorial
cohomologies describe in a unified manner gauge transformations,
fields and possible deformations of the models, \eg\ string-related
corrections in an $\a'$ expansion. Explicit cohomologies are
calculated for super-Yang--Mills theory in $D=10$, for the $\N=(2,0)$
tensor multiplet in $D=6$ and for supergravity
in $D=11$, in the latter case from the point of view of both 
the super-vielbein and the super-{\old3}-form potential. The techniques
may shed light on some questions concerning the $\a'$-corrected
effective theories, and result in better understanding of the r\^ole
of the {\old3}-form in $D=11$ supergravity.
\vfill

\line{\hrulefill}
\catcode`\@=11
\line{\tentt martin.cederwall@fy.chalmers.se\hfill}
\line{\tentt bengt.nilsson@fy.chalmers.se\hfill}
\line{\tentt tsimpis@fy.chalmers.se\hfill}
\catcode`\@=\active

\eject

\ref\CederwallNilssonTsimpisI{M. Cederwall, B.E.W. Nilsson and D. Tsimpis,
{\xit ``The structure of maximally supersymmetric gauge theories: 
constraining higher order interactions''}, 
\JHEP{01}{06}{2001}{034} [\hepth{0102009}].}

\ref\CederwallNilssonTsimpisII{M. Cederwall, B.E.W. Nilsson and D. Tsimpis,
{\xit ``D=10 super-Yang--Mills at $\ss O(\a'^2)$''}, 
\JHEP{01}{07}{2001}{042} [\hepth{0104236}].}

\ref\CederwallKarpacz{M. Cederwall, 
{\xit ``Superspace methods in string theory, supergravity and gauge theory''},
\hepth{0105176}.}

\ref\GatesNishinoII{S.J. Gates, Jr. and H. Nishino, 
{\xit ``Deliberations on 11D superspace for the M-theory effective action''},
\PLB{508}{2001}{155} [\hepth{0001037}].}

\ref\HoweWeyl{P.~Howe,
{\xit ``Weyl superspace''},
\PLB{415}{1997}{149} [\hepth{9707184}].}

\ref\NilssonSYM{B.E.W.~Nilsson, 
\xit ``Off-shell fields for the 10-dimensional supersymmetric 
Yang--Mills theory'', \xrm G\"oteborg-ITP-{\xold81}-{\xold6}.}

\ref\NilssonPure{B.E.W.~Nilsson, 
{\xit ``Pure spinors as auxiliary fields in the ten-dimensional 
supersymmetric Yang--Mills theory''},
\CQG3{1986}{{\xrm L}41}.}

\ref\LiE{A.M. Cohen, M. van Leeuwen and B. Lisser, 
LiE v. {\xold2}.{\xold2} ({\xold1998}), 
\nlni http://wallis.univ-poitiers.fr/\~{}maavl/LiE/} 

\ref\BergshoeffFFOUR{E.~Bergshoeff, M.~Rakowski and E.~Sezgin,
{\xit ``Higher derivative super Yang--Mills theories''},
\PLB{185}{1987}{371}.}

\ref\CGNN{M. Cederwall, U. Gran, M. Nielsen and B.E.W. Nilsson, 
{\xit ``Manifestly supersymmetric M-theory''}, 
\JHEP{00}{10}{2000}{041} [\hepth{0007035}];
{\xit ``Generalised 11-dimensional supergravity''}, \hepth{0010042}.}

\ref\CGNTprep{M. Cederwall, U. Gran, B.E.W. Nilsson and D. Tsimpis,
work in progress.}

\ref\SuperYM{L. Brink, J.H. Schwarz and J. Scherk, 
{\xit ``Supersymmetric Yang--Mills theories''},
\NPB{121}{1977}{77}.}

\ref\NilssonSixDSYM{B.E.W. Nilsson, 
{\xit ``Superspace action for a 6-dimensional non-extended supersymmetric
Yang--Mills theory''},
\NPB{174}{1980}{335}.}

\ref\GatesVashakidze{S.J. Gates, Jr. and Sh. Vashakidze,
{\xit ``On D=10, N=1 supersymmetry, superspace geometry 
and superstring effects''}, \PLB{226}{1989}{237}.}

\ref\HoweSezginWest{P.S. Howe, E. Sezgin and P.C. West,  
{\xit ``Aspects of superembeddings''},
\hepth{9705093}.}

\ref\SezginTopics{E. Sezgin, {\xit ``Topics in M-theory''},
\hepth{9809204}.}

\ref\NilssonBuckow{B.E.W.~Nilsson, {\xit ``A supersymmetric approach 
to branes and supergravity''}, 
in ``Theory of elementary particles'', Proc. of the
{\xold31}st international symposium Ahrenshoop, September {\xold2}-{\xold6}, 
{\xold1997}, Buckow,
Eds H. Dorn et al. (Wiley-VCH {\xold1998}) [\hepth{0007017}].}

\ref\ElevenSG{E. Cremmer, B. Julia and J. Sherk, 
{\xit ``Supergravity theory in eleven-dimensions''},
\PLB{76}{1978}{409}.}

\ref\ElevenSSSG{L. Brink and P. Howe, 
{\xit ``Eleven-dimensional supergravity on the mass-shell in superspace''},
\PLB{91}{1980}{384};\nlni
E. Cremmer and S. Ferrara,
{\xit ``Formulation of eleven-dimensional supergravity in superspace''},
\PLB{91}{1980}{61}.}

\ref\SuperYM{L. Brink, J.H. Schwarz and J. Scherk, 
{\xit ``Supersymmetric Yang--Mills theories''},
\NPB{121}{1977}{77}.}

\ref\Kappa{E. Bergshoeff, E. Sezgin and P.K. Townsend, 
{\xit ``Supermembranes and eleven-dimensional supergravity''},
\PLB{189}{1987}{75};
{\xit ``Properties of the eleven-dimensional supermembrane theory''},
\AP{185}{1988}{330};\nlni
M.J. Duff, P.S. Howe, T. Inami and K.S. Stelle,
{\xit ``Superstrings in D=10 from supermembranes in D=11''},
\PLB{191}{1987}{70};\nlni
M. Cederwall, B.E.W. Nilsson and P. Sundell,
{\xit ``An action for the super-5-brane in D=11 supergravity''},
\JHEP{98}{04}{1998}{007} [\hepth{9712059}];\nlni
M. Cederwall, A. von Gussich, B.E.W. Nilsson and A. Westerberg,
{\xit ``The Dirichlet super-three-brane in ten-dimensional type IIB 
supergravity''}
\NPB{490}{1997}{163} [\hepth{9610148}];\nlni
M. Aganagi\'c, C. Popescu, J.H. Schwarz,
{\xit ``D-brane actions with local kappa symmetry''},
\PLB{393}{1997}{311} [\hepth{9610249}];\nlni
M. Cederwall, A. von Gussich, B.E.W. Nilsson, P. Sundell
 and A. Westerberg,
{\xit ``The Dirichlet super-p-branes in ten-dimensional type IIA and IIB 
supergravity''},
\NPB{490}{1997}{179} [\hepth{9611159}];\nlni
E. Bergshoeff and P.K. Townsend, 
{\xit ``Super D-branes''},
\NPB{490}{1997}{145} [\hepth{9611173}].}

\ref\BergshoeffDeRoo{E. Bergshoeff and M. de Roo,
{\xit ``The supercurrent in ten dimensions''},
\PLB{112}{1982}{53}.}

\ref\GAMMA{U. Gran,
{\xit ``GAMMA: A Mathematica package for performing gamma-matrix 
algebra and Fierz transformations in arbitrary dimensions''},
\hepth{0105086}.}

\ref\PVW{K. Peeters, P. Vanhove and A. Westerberg,
{\xit ``Supersymmetric higher derivative actions in ten-dimensions 
and eleven-dimensions,
the associated superalgebras and their formulation in superspace''},
\CQG{18}{2001}{843} [\hepth{0010167}];
{\xit ``Supersymmetric R${}^4$ actions and quantum corrections 
to superspace torsion constraints''},
\hepth{0010182}.}

\ref\HowePrivate{P. Howe, private communication.}

\ref\BerkovitsCohomology{N. Berkovits,
{\xit ``Cohomology in the pure spinor formalism for the superstring''},
\jhep{00}{09}{2000}{046} [\hepth{0006003}].}

\section\Intro{Introduction}The purpose of this paper is to investigate
an interesting structure of supersymmetric field theories, recently
found when deriving conditions on interactions in maximally supersymmetric
Yang--Mills theory 
[\CederwallNilssonTsimpisI,\CederwallNilssonTsimpisII,\CederwallKarpacz]. 
It was observed that
component fields and gauge transformations, 
as well as physically distinguishable deformations of
the model, are represented as elements in cohomology classes under a
certain fermionic exterior derivative. Since the model under consideration,
super-Yang--Mills theory in $D=10$,
is a maximally supersymmetric model, known to possess no manifestly
supersymmetric off-shell formulation, the deformations are represented
(at least at the linearised level) as a current supermultiplet.
In addition to promoting a better understanding of the mechanisms
at hand for such theories, the concept turned out to be quite efficient
in understanding field redefinitions relating physically equivalent
deformations.

The structure we are dealing with is a sequence of representations of
the global symmetry, which is the Lorentz group $L$
(together with the $R$-symmetry
group $R$ if there is one---most of the cases we deal with in the present paper
have trivial $R$-symmetry). The examples we will
treat are 10-dimensional super-Yang--Mills theory [\SuperYM],
the $\N=(2,0)$ tensor multiplet in $D=6$ 
and 11-dimensional supergravity [\ElevenSG,\ElevenSSSG], 
the latter one both from the perspective of the vielbein
and from that of the three-form tensor. 
The treatment of dimensional reductions 
of these theories will follow from
dimensional reduction of the complexes.
Another case, which we expect to show similar properties, is type \II B
supergravity in $D=10$, which we will not treat in this paper.
$\N=(1,0)$ super-Yang--Mills in $D=6$ will also be analysed, as
a contrasting example of a theory possessing a
supersymmetric off-shell formulation.

The basic idea is that the theories we consider are gauge theories, and
that, in a superspace formulation, where all potentials and field strengths
are forms on superspace, all components except the purely spinorial ones
are redundant. Since all physical fields are contained in the objects 
carrying spinorial form indices only, it is interesting to examine the
structure arising from these.
Our complexes are of the form
$$
r_{0}\Darrow0r_{1}\Darrow1r_{2}\Darrow2\ldots\Darrow{n-1}r_{n}\Darrow n\ldots
\komma\Eqn\Basic
$$
where $r_{p}$, for some $p\geq0$, 
is the representation carried by a gauge transformation,
$r_{p+1}$ that of a potential and $r_{p+2}$ that of a field strength.
We will refer to the representations $r_{n}$ as $n$-forms, a notation
not to be confused with that of a tensor antisymmetric in vector indices.
The exact definitions are given, both for gauge theory and supergravity,
in the following sections, where it will also be clear why $\D$
is a nilpotent operator. The r\^ole of $r_{p+3}$ is as a Bianchi identity.

A supersymmetric gauge theory or a supergravity theory, when formulated
on superspace, has to fulfill a number of Bianchi identities. These are
of course trivial as long as field strengths are defined from potentials,
but become non-trivial when conventional constraints are imposed on
the field strengths. Then one can use the Bianchi identities as integrability
conditions and work only at the level of field strengths. For the maximally
supersymmetric theories we are dealing with, this procedure enforces
equations of motion for the component fields, which however depend on
the choice of the fermionic components of the field strengths, which 
in a certain sense, explained in the following, play the r\^oles of 
current multiplet superfields. 
Since the structure we are presenting only deals with
the totally fermionic forms, we ignore all Bianchi identities except
the spinorial one. We are thus not performing a complete analysis of the 
theories, for which the reader should consult other references 
[\CederwallNilssonTsimpisI,\CederwallNilssonTsimpisII,\HoweWeyl,\CGNN,\CGNTprep,\PVW]. 
The analysis of this paper
demonstrates, among other things, the {\it possibility} of deforming the 
theories in a non-trivial manner by turning on field strengths. That
such a procedure is consistent follows from a full analysis of the
Bianchi identities, which will also yield the exact form of the deformations.
This has been done for $D=10$ super-Yang--Mills
[\CederwallNilssonTsimpisI,\CederwallNilssonTsimpisII], while for
$D=11$ supergravity a partial analysis has been performed [\CGNN] and
a more complete one is envisaged [\CGNTprep].

\section\Cohomology{Spinorial complexes and cohomology}The 
structure of the complex is 
$$
r_{0}\Darrow0r_{1}\Darrow1r_{2}\Darrow2\ldots\Darrow{n-1}r_{n}\Darrow n\ldots
\Eqn\BasicComplex
$$
where the representations $r_{n}$ at each $n$ denotes a superfield in
the representation $r_{n}$ of the Lorentz group. When we describe
a gauge theory, $r_{n}$ consists of
totally symmetric and $\G$-traceless 
tensors in $n$ spinor indices. The fermionic
exterior derivative is a projection on the representations $r_{n}$ of
a (symmetrised) spinorial covariant derivative. The general
interpretation is that $r_{0}$ contains gauge transformations, $r_{1}$
contains fields and $r_{2}$ deformations of the theory.  
For a theory of a rank-$(p+1)$ tensor potential, such as the three-form
in $D=11$ supergravity, $r_{p}$ contains gauge transformations, $r_{p+1}$
fields and $r_{p+2}$ deformations.
The representations in a complex associated with the vielbein of a 
supergravity theory are slightly different. Then $r_{0}$ is a vector,
corresponding to reparametrisation gauge transformations, $r_1$ is
a vector-spinor, and so on.

It is convenient to decompose the representations in component fields
sitting at levels $\ell$, \ie, multiplying $\theta^\ell$ in the
superfields.  We write those as $r_{n}^\ell\equiv\wedge^\ell S\otimes
r_{n}$, where $S$ is the representation of the spinorial derivative
(\ie, the conjugate representation to that of $\theta$).  
Appendix B gives, as an example, all
antisymmetric products of chiral spinors in $D=10$ [\BergshoeffDeRoo].  

We will argue from very general arguments that the  
interesting content at each $n$ actually is the cohomology with respect
to the exterior derivative $\D$, $\H^n=\hbox{Ker}\D_n/\hbox{Im}\D_{n-1}$.
We shall give a couple of examples of this for maximally supersymmetric
theories, and also compare to the situation in theories with 
lower supersymmetry.

We use the Dynkin labels of highest weights to denote irreducible
representations of $L\times R$.
The $D=10$ super-Yang--Mills complex is:
$$
(00000)\Darrow0(00010)\Darrow1(00020)\Darrow2\ldots\Darrow{n-1}(000n0)
\Darrow n\ldots\Eqn\TenSYMComplex
$$
The complex for the $\N=(2,0)$ tensor multiplet in $D=6$, with
$R=\hbox{Sp}(4)$, is
$$
\vtop{\baselineskip0pt\lineskip0pt
\ialign{
$\hfill#\hfill$&$\,\hfill#\hfill\,$&$\hfill#\hfill$&$\,\hfill#\hfill\,$&$\hfill#\hfill$&$\,\hfill#\hfill\,$&$\hfill#\hfill$&$\,\hfill#\hfill\,$&$\hfill#\hfill$&$\,\hfill#\hfill\,$&$\hfill#\hfill$&$\hfill#\hfill$\cr
(000)(00)&\ra&(100)(10)&\ra &(200)(20)&\ra &(300)(30)&\ra &(400)(40)&
								\ldots\cr
       &   &       &\sea&       &\sea&       &\sea&       &&&\cr
       &   &       &    &(010)(01)&\ra &(110)(11)&\ra &(210)(21)&
								\ldots\cr
       &   &       &    &       &    &       &\sea&       &&&      \cr
       &   &       &    &       &    &       &    &(020)(02)&
								\ldots\cr
       &   &       &    &       &    &       &    &       &&&\phantom{\ra}\cr
}}\Eqn\SixTensorComplex
$$
The corresponding complex in $D=11$, which we will apply to the {\old3}-form
present in $D=11$ supergravity, is 
$$
\vtop{\baselineskip0pt\lineskip0pt
\ialign{
$\hfill#\hfill$&$\,\hfill#\hfill\,$&$\hfill#\hfill$&$\,\hfill#\hfill\,$&$\hfill#\hfill$&$\,\hfill#\hfill\,$&$\hfill#\hfill$&$\,\hfill#\hfill\,$&$\hfill#\hfill$&$\,\hfill#\hfill\,$&$\hfill#\hfill$&$\hfill#\hfill$\cr
(00000)&\ra&(00001)&\ra &(00002)&\ra &(00003)&\ra &(00004)&\ra &(00005)&
								\ldots\cr
       &   &       &\sea&       &\sea&       &\sea&       &\sea&&\cr
       &   &       &    &(01000)&\ra &(01001)&\ra &(01002)&\ra &(01003)&
								\ldots\cr
       &   &       &    &       &    &       &\sea&       &\sea&&      \cr
       &   &       &    &       &    &       &    &(02000)&\ra &(02001)&
								\ldots\cr
       &   &       &    &       &    &       &    &       &&&\phantom{\ra}\cr
}}\Eqn\ElevenComplex
$$
The $D=11$ supergravity complex, which is obtained from the last example
by adding the vector highest weight (10000), is:
$$
\vtop{\baselineskip0pt\lineskip0pt
\ialign{
$\hfill#\hfill$&$\,\hfill#\hfill\,$&$\hfill#\hfill$&$\,\hfill#\hfill\,$&$\hfill#\hfill$&$\,\hfill#\hfill\,$&$\hfill#\hfill$&$\,\hfill#\hfill\,$&$\hfill#\hfill$&$\hfill#\hfill$\cr
(10000)&\ra&(10001)&\ra &(10002)&\ra &(10003)&\ra &(10004)&\ldots\cr
       &   &       &\sea&       &\sea&       &\sea&       &      \cr
       &   &       &    &(11000)&\ra &(11001)&\ra &(11002)&\ldots\cr
       &   &       &    &       &    &       &\sea&       &      \cr
       &   &       &    &       &    &       &    &(12000)&\ldots\cr
       &   &       &    &       &    &       &    &       &\phantom{\ra}\cr
}}\Eqn\ElevenSGComplex
$$
In contrast to the one for $D=10$ SYM, the $D=11$ complexes contain reducible
representations for $n\geq2$, simply because the symmetric bi-spinors,
apart from the vector, decompose into an {\old2}-form, (01000), and
a {\old5}-form, (00002).
The situation in $D=6$ is similar; here the symmetric bispinors contain
a vector which is an Sp(4) singlet, (010)(00), a vector in {\bf5} of Sp(4),
(010)(01), and a self-dual {\old3}-form in {\bf10} of Sp(4), (200)(20).

Here we have not bothered to
name the operators taking us between irreducible representations,
only indicated with arrows which paths are possible.  

Let us describe in more detail how the complexes work, with the
super-Yang--Mills theory as an example. 
The gauge potentials are $A_\a$ and $A_a$. However, the spinor potential
already contains a vector (of correct dimension) at the $\theta$ level,
and this is the reason why a conventional constraint is needed in order
to have {\it one} vector potential. This constraint is 
$$
\G_a^{\a\b}F_{\a\b}=0\komma\eqn
$$
which implies that 
$$
A_a=-\fr{16}\left(D\G_aA-A\G_aA\right)\punkt\eqn
$$
The rest of $F_{\a\b}$, 
$$
F_{\a\b}=\fr{5!}\G_{\a\b}^{a_1\ldots a_5}J_{a_1\ldots a_5}\komma\eqn
$$
which lies in (00020), does not contain $A_a$.
We also note [\NilssonSYM,\NilssonPure,\CederwallNilssonTsimpisI] 
that part of the dimension-$\Fr32$ Bianchi identity
states the vanishing of the (00030) component of $D_\a F_{\b\g}$.
These observations make it natural to consider, 
not the sequence of completely symmetric
representations in spinor indices, but a restriction of it, namely 
the sequence of Spin(1,9) representations in eq. (\TenSYMComplex).
The representation $r_n\equiv(000n0)$ is the part of the totally symmetric
product of $n$ chiral spinors that has vanishing ``$\G$-trace'', and may
be represented tensorially as 
$C_{\a_1\ldots\a_n}=C_{(\a_1\ldots\a_n)}$, 
$\G_a{}^{\a_1\a_2}C_{\a_1\a_2\a_3\ldots\a_n}=0$.
For $n=2$, $C$ is an anti-selfdual five-form, for $n=3$ a $\G$-traceless
anti-selfdual five-form spinor, etc.

The operator $\D_n$: $r_n\lra r_{n+1}$ can schematically be written
as $\D_nC_n=\Pi(r_{n+1})DC_n$, where $D$ is the exterior covariant derivative
$D=d\theta^\a D_\a$ and $\Pi(r_n)$ is the algebraic projection
from $\otimes^n_s(00010)$ to $(000n0)$. It is straightforward to write
an explicit tensorial form for $\D$ by subtracting $\G$-traces from $DC$,
but it will not be used here.
It is also straightforward to show that, for an abelian gauge group
and standard flat superspace,
the sequence (\TenSYMComplex) forms
a complex, \ie, that $\D^2=0$. This follows simply from the fact that
while $\{D_\a,D_\b\}=-T_{\a\b}{}^cD_c$, the torsion only has a component
$2\G_{\a\b}{}^c$ which is projected out by $\Pi(r_n)$.
The anticommutator of two covariant spinor derivatives is in general
$$
\{D_\a,D_\b\}=-T_{\a\b}{}^cD_c-T_{\a\b}{}^\g D_\g
	+F_{\a\b}\!\cdot+R_{\a\b}\!\cdot\komma\Eqn\GeneralAC
$$
and $\D$ will not be nilpotent in arbitrary curved backgrounds or non-abelian
theories. In these cases, we must consider the complex for an undeformed
super-Yang--Mills or supergravity theory, and consider 
{\it infinitesimal} deformations as elements of the cohomologies.
That eq. (\GeneralAC) in the undeformed theories yield $\D^2=0$ is seen
as follows. For the Yang--Mills case, $F_{\a\b}=0$ in the undeformed theory.
For $D=11$ supergravity the argument for $\D^2=0$ in the undeformed
theory is slightly more complicated. One has to remember that $T_{a\b}{}^\g$ is
non-zero (it contains the {\old4}-form tensor field strength $H$).
The torsion Bianchi identities at dimension 1 give
$$
\eqalign{
R_{(\a\b\g)}{}^\d&=6\G_{(\a\b}{}^eT_{|e|\g)}{}^\d\komma\cr
R_{\a\b c}{}^d&=-4T_{c(\a}{}^\g\G_{\b)\g}{}^d\punkt\cr
}\Eqn\DimOneTBI
$$
Letting two consecutive spinorial derivatives act on an element
$C_{\g_1\ldots\g_n}$ in the sequence (\ElevenComplex) or 
$C_{\g_1\ldots\g_n}{}^c$ in (\ElevenSGComplex) gives additional curvature
terms according to eq. (\GeneralAC).
Inserting the expressions for the dimension-1 curvature of eq. 
(\DimOneTBI) implies that the resulting expressions can be written as
$$
\eqalign{
D_{(\a}D_{\mathstrut\b}C_{\g_1\ldots\g_n)}
	&=\G_{(\a\b}{}^d\phi_{|d|\g_1\ldots\g_n)}
	\komma\cr
D_{(\a}D_{\mathstrut\b}C_{\g_1\ldots\g_n)}{}^c
	&=\G_{(\a\b}{}^d\psi_{|d|\g_1\ldots\g_n)}{}^c
	+\G^c{}_{(\a}{}^\d\chi_{|\d|\b\g_1\ldots\g_n)}\punkt\cr
}\eqn
$$
Each of these terms vanish under the projection on the irreducible
representations constituting the complexes---while the representations
$r_n$ are ``$\G$-traceless'', they contain pure $\G$-traces only.

We would now like to calculate the cohomology of the complex associated
with $D=10$ super-Yang--Mills. This can be done by considering
the decomposition into irreducible representations of the representation
sitting at level $\ell$ in $r_n$, $r_{n}^\ell\equiv\wedge^\ell S\otimes
r_{n}$. This is easily done, \eg\ with the help of the program LiE [\LiE].
One then follows each of the irreducible representations at a given
dimension through the subcomplex
$$
r_0^\ell\ra r_1^{\ell-1}
\ra r_2^{\ell-2}
\ra\ldots\ra r_{\ell-1}^{1}\ra
r_\ell\punkt\Eqn\Subcomplex
$$

Let us illustrate the calculation by examining the field content.
We then look into the spinor potential of dimension $\half$, 
which contains all fields
in the vector multiplet, so we should examine the first cohomology.
The vector (dimension 1) sits at
$\ell=\half$ and the spinor (dimension $\Fr32$) at $\ell=1$. 
The subcomplexes under consideration are 
$r_0^2\ra r_1^1\ra r_2$
and 
$r_0^3\ra r_1^2\ra r_2^1\ra r_3$.
Checking the multiplicities of the relevant representations, (10000) and
(00001), in these, we obtain the sequences
$0\ra1\ra0$ and $0\ra1\ra0\ra0$. The components of the cohomology
in these representations and dimensions clearly contain the physical
fields. This can be understood in a traditional framework as removing
degrees of freedom in a superfield gauge transformation (removing
the image from the left) and imposing the vanishing of the field
strength $F_{\a\b}$ (removing the complement of the kernel from the
right).
Analogous considerations tell us that the second cohomology contains
a spinor of dimension $\Fr52$ and a vector of dimension $3$. These are
interpreted as belonging to a current supermultiplet, \ie, fields
entering the right hand sides of the equations of motion.
This goes well together with the observation that modifications of
the theory are introduced by deforming the constraint $F_{\a\b}=0$
[\NilssonSYM,\BergshoeffFFOUR,\GatesVashakidze,\CederwallNilssonTsimpisI,\CederwallNilssonTsimpisII].
The relevance of the cohomology is explained by the facts that deformations
introduced by relaxing $F_{\a\b}=0$ have to fulfill the Bianchi identity
(removing the complement of the kernel from the right),
and that relevant deformations are counted modulo field redefinitions
(removing the image from the left).
See also the following section for a fuller discussion.

A complete calculation of the cohomology requires that one considers
all irreducible representations occurring at arbitrary levels.
This quickly becomes untractable to do by hand. Unfortunately, there
is also another complication, that makes it impossible to derive
the cohomologies unambiguously from multiplicities only without making
further assumptions. This can be exemplified by looking at some
other representation; let us take the {\old3}-form at dimension 3.
The subcomplex 
$r_0^6\ra r_1^5\ra r_2^4\ra r_3^3\ra r_4^2\ra r_5^1\ra r_6$ 
yields the multiplicities
$0\ra0\ra1\ra1\ra0\ra0\ra0$. We do not expect any non-zero cohomology,
since there is no equation of motion in this representation.
Yet, the sequence of multiplicities offers two possibilities:
either there is one {\old3}-form in the first cohomology and one in
the second, or there is none at all. Using tensorial methods, it is
easy to show that the {\old3}-form in $r_2^4$ has an image in $r_3^3$,
so the cohomology vanishes.
In the present paper, we will make the assumption of 
``maximal propagation'' of irreducible representations through the
subcomplexes, meaning that representations have images or belong to
images under $\Delta$ if possible. This assumption is enough to
determine the super-Yang--Mills cohomology completely, at least for $n\leq5$.
The result, derived already in ref. [\CederwallNilssonTsimpisI], 
is presented in table {\old1}. We expect higher cohomologies to vanish.
The method for calculating cohomologies, under the assumption of maximal
propagation, is by using the code of appendix A with the program LiE [\LiE].
As we will see, there are cases in $D=11$ where even the 
assumptions made so far
leave an ambiguity. The reason that we choose to make educated guesses
rather than turn to tensor calculations is that the number of 
irreducible representations is so large that such a treatment becomes
virtually impossible.

We now turn to the cohomology of the complex (\ElevenSGComplex) 
associated with the super-vielbein of $D=11$ supergravity. 
All fields in the supergravity multiplet are contained in the 
dimen\-sion-$(-\half)$ vielbein $E_\a{}^a$
(actually in the $\G$-traceless part (10001)),
which plays an analogous r\^ole to that of $A_\a$ in super-Yang--Mills theory.
All other components should be related to this one by conventional
constraints. 
When one considers the corresponding field strength, the dimension-0 torsion
component $T_{\a\b}{}^a$, it is known 
[\NilssonBuckow,\HoweSezginWest,\SezginTopics,\CGNN] that the only components
surviving after imposing conventional constraints are the usual 
constant $\G$-matrix term, and two fields 
$X_{a_1a_2}{}^a$ in (11000) and $X_{a_1\ldots a_5}{}^a$ in (10002)
entering the torsion as
$$
T_{\a\b}{}^a=2\left(\G_{\a\b}^a+\half\G_{\a\b}^{a_1a_2}X_{a_1a_2}{}^a
+\fr{5!}\G_{\a\b}^{a_1\ldots a_5}X_{a_1\ldots a_5}{}^a\right)\punkt\eqn
$$
The part of the torsion Bianchi identity with purely spinorial 
form indices is
$$
D_{(\a}T_{\b\g)}{}^a|_{(11001)\oplus(10003)}=0\eqn
$$
(analagously to eq. (\GeneralAC), there are higher order torsion terms 
that should be added to this equation, 
which do not contribute to the relevant representations
for infinitesimal deformations).
We are naturally led to consider the sequence of representations
already stated in eq. (\ElevenSGComplex).

The cohomology
is given in table {\old3}, but we would like to illustrate part of the
calculation. The representations in the zeroth and first cohomologies
are calculated the same way as for super-Yang--Mills.
When we come to the second cohomology, representing the current
supermultiplet contained in the torsion components (11000) and (10002)
at dimension 0 [\CGNN], there is a complication illustrated by the
following example. 
Take the spinor (00001) at dimension $3\over2$.  It will occur in the
equation of motion for the gravitino, if it is contained in the $n=2$
cohomology (that this is the case, and that it does not affect the
Weyl curvature at dimension $3\over2$, was actually shown in ref. 
[\CGNN]).  We now only write the multiplicity of the representation in
each $r_{n}^\ell$, and get the sequence
$1\ra3\ra3\ra0\ra0\ra0$. This sequence offers two distinct possibilities
even under the assumption of maximal propagation:
either the representation in $r_0$ has an image among the three in
$r_1$, in which case there is a cohomology in $r_2$, or all three
representations in $r_2$ have images in $r_3$, in which case there
is a cohomology in $r_0$. In this specific case, tensorial methods
have already been used [\CGNN] that show that the first of these
possibilities is true, so that the second cohomology contains a spinor
in the equations of motion.
The r\^ole of the code given in appendix A is that it makes use
of the assumption of maximal propagation, and in cases like the
one just related, gives candidate cohomologies in all possible
cases. For the cohomologies associated with the $D=11$ super-vielbein
and super-{\old3}-form, such ambiguities exist when one goes higher
in dimension than those of the fields, and the results of tables {\old3} 
and {\old4} consist partially of educated guesses concerning which alternatives
are the correct ones. We are led in part by expectations concerning
the content of the current supermultiplet and in part by the resulting
symmetry of the tables---all three cohomologies in 10 and 11 dimensions
seem to have an
inherent symmetry under reflection in one point (in the $D=10$ case
accompanied by a ${\Bbb Z}_2$ automorphism exchanging the
spinor representations of Spin(1,9)).

To summarise the calculation of the cohomologies,
the non-vanishing cohomology associated to $D=10$ super-Yang--Mills theory is
$$
\eqalign{
\H^0&=(00000)_0\komma\cr
\H^1&=(10000)_1\oplus(00001)_{3\over2}\komma\cr
\H^2&=(00010)_{5\over2}\oplus(10000)_3\komma\cr
\H^3&=(00000)_4\cr
}\eqn
$$
(the dimensions are given as subscripts),
or represented graphically in a table, divided 
in different $n$ and $\ell$ (the dimension is ${n+\ell\over2}$):

\vskip2\parskip
\pagegoal=30cm

$$
\vtop{\baselineskip20pt\lineskip0pt
\ialign{
$\hfill#\quad$&$\,\hfill#\hfill\,$&$\,\hfill#\hfill\,$&$\,\hfill#\hfill\,$
&$\,\hfill#\hfill\,$&$\,\hfill#\hfill$&\qquad\qquad#\cr
            &n=0    &n=1    &n=2    &n=3    &n=4\cr
\hbox{dim}=0&(00000)&       &       &       &\phantom{(00000)}       \cr
        \fr2&\bullet&\bullet&               &       \cr 
           1&\bullet&(10000)&\bullet&       &       \cr
       \Fr32&\bullet&(00001)&\bullet&\bullet&       \cr
           2&\bullet&\bullet&\bullet&\bullet&\bullet\cr
       \Fr52&\bullet&\bullet&(00010)&\bullet&\bullet\cr
           3&\bullet&\bullet&(10000)&\bullet&\bullet\cr
       \Fr72&\bullet&\bullet&\bullet&\bullet&\bullet\cr
           4&\bullet&\bullet&\bullet&(00000)&\bullet\cr
       \Fr92&\bullet&\bullet&\bullet&\bullet&\bullet\cr
}}
$$

\vskip2\parskip
\centerline{\it Table 1. The cohomology of the $D=10$ SYM complex.}

\vfill\eject

The cohomology for the $\N=(2,0)$ tensor multiplet in $D=6$ and
the $D=11$ cohomologies for the super-vielbein and the tensor
are given in the following three tables:

\vskip2\parskip

$$
\vtop{\baselineskip20pt\lineskip0pt
\ialign{
$\hfill#\quad$&$\,\hfill#\hfill\,$&$\,\hfill#\hfill\,$&$\,\hfill#\hfill\,$
&$\,\hfill#\hfill\,$&$\,\hfill#\hfill$&\qquad\qquad#\cr
            &n=0    &n=1    &n=2    &n=3    &n=4\cr
\hbox{dim}=0&(000)(00)&       &       &       &\phantom{(000)(00)}       \cr
        \fr2&\bullet&\bullet&               &       \cr 
           1&\bullet&(010)(00)&\bullet&       &       \cr
       \Fr32&\bullet&\bullet&\bullet&\bullet&       \cr
           2&\bullet&\bullet&\raise5pt\vtop{\baselineskip6pt\ialign{
					\hfill$#$\hfill\cr
					(000)(01)\cr
					(101)(00)\cr}}&\bullet&\bullet\cr
       \Fr52&\bullet&\bullet&(100)(10)&\bullet&\bullet\cr
           3&\bullet&\bullet&\bullet&(002)(00)&\bullet\cr
       \Fr72&\bullet&\bullet&\bullet&(001)(10)&\bullet\cr
           4&\bullet&\bullet&\bullet&(000)(01)&\bullet\cr
       \Fr92&\bullet&\bullet&\bullet&\bullet&\bullet\cr
}}
$$

\vskip2\parskip
\centerline{\it Table 2. The cohomology of the $D=6$, $\N=(2,0)$ complex.}

\vfill\eject

$$\hskip-2cm
\vtop{\baselineskip25pt\lineskip0pt
\ialign{
$\hfill#\quad$&$\ss\,\hfill#\hfill\,$&$\ss\,\hfill#\hfill\,$
&$\ss\,\hfill#\hfill\,$&$\ss\,\hfill#\hfill\,$&$\ss\,\hfill#\hfill$
&$\ss\,\hfill#\hfill$&$\ss\,\hfill#\hfill$
		&\quad#\cr
            &\ts n=0&\ts n=1&\ts n=2&\ts n=3&\ts n=4&\ts n=5&\ts n=6&\cr
\hbox{dim}=-1&\quad(10000)\quad
		&\quad\phantom{(00000)}\quad
		&\quad\phantom{(00000)}\quad
		&\quad\phantom{(00000)}\quad
		&\quad\phantom{(00000)}\quad
		&\quad\phantom{(00000)}\quad
		&\quad\phantom{(00000)}\quad
		&\cr
        -\fr2&(00001)&\bullet&               &&&       &\cr 
           0&\bullet&(20000)&\bullet&       &&&       &\cr
       \Fr12&\bullet&\raise3pt\vtop{\baselineskip6pt\ialign{
					\hfill$#$\hfill\cr
					\ss(00001)\cr
					\ss(10001)\cr}}
			&\bullet&\bullet&&&&\cr
           1&\bullet&\raise3pt\vtop{\baselineskip6pt\ialign{
					\hfill$#$\hfill\cr
					\ss(00010)\cr
					\ss(10000)\cr}}
			&\bullet&\bullet&\bullet&&&\cr
       \Fr32&\bullet&\bullet&\raise3pt\vtop{\baselineskip6pt\ialign{
					\hfill$#$\hfill\cr
					\ss(00001)\cr
					\ss(10001)\cr}}
			&\bullet&\bullet&\bullet&&\cr
           2&\bullet&\bullet&\raise6pt\vtop{\baselineskip6pt\ialign{
					\hfill$#$\hfill\cr
					\ss(00000)(00002)\cr
					\ss(00100)(01000)\cr
					\ss(10000)(20000)\cr}}
				&\bullet&\bullet&\bullet&\bullet&\cr
       \Fr52&\bullet&\bullet&\bullet&\bullet&\bullet&\bullet&\bullet&\cr
           3&\bullet&\bullet&\bullet&\raise6pt\vtop{\baselineskip6pt\ialign{
					\hfill$#$\hfill\cr
					\ss(00000)(00002)\cr
					\ss(00100)(01000)\cr
					\ss(10000)(20000)\cr}}
				&\bullet&\bullet&\bullet\cr
       \Fr72&\bullet&\bullet&\bullet&\raise3pt\vtop{\baselineskip6pt\ialign{
					\hfill$#$\hfill\cr
					\ss(00001)\cr
					\ss(10001)\cr}}
			&\bullet&\bullet&\bullet&\cr
           4&\bullet&\bullet&\bullet&\bullet
				&\raise3pt\vtop{\baselineskip6pt\ialign{
					\hfill$#$\hfill\cr
					\ss(00010)\cr
					\ss(10000)\cr}}
			&\bullet&\bullet&\cr
       \Fr92&\bullet&\bullet&\bullet&\bullet
				&\raise3pt\vtop{\baselineskip6pt\ialign{
					\hfill$#$\hfill\cr
					\ss(00001)\cr
					\ss(10001)\cr}}
			&\bullet&\bullet&\cr
       	   5&\bullet&\bullet&\bullet&\bullet&(20000)&\bullet&\bullet&\cr
    \Fr{11}2&\bullet&\bullet&\bullet&\bullet&\bullet&(00001)&\bullet&\cr
       	   6&\bullet&\bullet&\bullet&\bullet&\bullet&(10000)&\bullet&\cr
       \Fr{13}2&\bullet&\bullet&\bullet&\bullet&\bullet&\bullet&\bullet&\cr
}}
$$

\noindent{\it Table 3. The $D=11$ supergravity cohomology, with respect to
the super-vielbein.}


$$\hskip-2cm
\vtop{\baselineskip25pt\lineskip0pt
\ialign{
$\hfill#\quad$&$\ss\,\hfill#\hfill\,$&$\ss\,\hfill#\hfill\,$
&$\ss\,\hfill#\hfill\,$&$\ss\,\hfill#\hfill\,$&$\ss\,\hfill#\hfill$
&$\ss\,\hfill#\hfill$&$\ss\,\hfill#\hfill$
&$\ss\,\hfill#\hfill$&$\ss\,\hfill#\hfill$&\quad#\cr
&\ts n=0&\ts n=1&\ts n=2&\ts n=3&\ts n=4&\ts n=5&\ts n=6&\ts n=7&\ts n=8&\cr
\hbox{dim}=-3&\,\,(00000)\,\,
		&\phantom{\,\,(00000)\,\,}&\phantom{\,\,(00000)\,\,}
		&\phantom{\,\,(00000)\,\,}&\phantom{\,\,(00000)\,\,}
		&\phantom{\,\,(00000)\,\,}&\phantom{\,\,(00000)\,\,}
		&\phantom{\,\,(00000)\,\,}&\phantom{\,\,(00000)\,\,}&\cr
        -\Fr52&\bullet&\bullet&               &&&       &\cr 
           -2&\bullet&(10000)&\bullet&       &&&       &\cr
       -\Fr32&\bullet&\bullet&\bullet&\bullet&&&       &\cr
           -1&\bullet&\bullet&\raise3pt\vtop{\baselineskip6pt\ialign{
					\hfill$#$\hfill\cr
					\ss(01000)\cr
					\ss(10000)\cr}}
			&\bullet&\bullet&&\cr
       -\Fr12&\bullet&\bullet&(00001)
				&\bullet&\bullet&\bullet&&\cr
           0&\bullet&\bullet&\bullet&\raise6pt\vtop{\baselineskip6pt\ialign{
					\hfill$#$\hfill\cr
					\ss(00000)\cr
					\ss(00100)\cr
					\ss(20000)\cr}}
				&\bullet&\bullet&\bullet&&\cr
       \Fr12&\bullet&\bullet&\bullet&\raise3pt\vtop{\baselineskip6pt\ialign{
					\hfill$#$\hfill\cr
					\ss(00001)\cr
					\ss(10001)\cr}}
				&\bullet&\bullet&\bullet&\bullet&\cr
           1&\bullet&\bullet&\bullet&\bullet&\bullet&\bullet
			&\bullet&\bullet&\bullet&\cr
       \Fr32&\bullet&\bullet&\bullet&\bullet
				&\raise3pt\vtop{\baselineskip6pt\ialign{
					\hfill$#$\hfill\cr
					\ss(00001)\cr
					\ss(10001)\cr}}
				&\bullet&\bullet&\bullet&\bullet\cr
           2&\bullet&\bullet&\bullet&\bullet
				&\raise6pt\vtop{\baselineskip6pt\ialign{
					\hfill$#$\hfill\cr
					\ss(00000)\cr
					\ss(00100)\cr
					\ss(20000)\cr}}
				&\bullet&\bullet&\bullet&\bullet&\cr
       \Fr52&\bullet&\bullet&\bullet&\bullet&\bullet&(00001)&\bullet
			&\bullet&\bullet&\cr
       	   3&\bullet&\bullet&\bullet&\bullet&\bullet
				&\raise3pt\vtop{\baselineskip6pt\ialign{
					\hfill$#$\hfill\cr
					\ss(01000)\cr
					\ss(10000)\cr}}
				&\bullet&\bullet&\bullet&\cr
       \Fr72&\bullet&\bullet&\bullet&\bullet&\bullet&\bullet
				&\bullet&\bullet&\bullet&\cr
       	   4&\bullet&\bullet&\bullet&\bullet&\bullet&\bullet&(10000)
			&\bullet&\bullet&\cr
       \Fr92&\bullet&\bullet&\bullet&\bullet&\bullet&\bullet&\bullet
			&\bullet&\bullet&\cr
       	   5&\bullet&\bullet&\bullet&\bullet&\bullet&\bullet&\bullet
				&(00000)&\bullet&\cr
    \Fr{11}2&\bullet&\bullet&\bullet&\bullet&\bullet&\bullet&\bullet
			&\bullet&\bullet&\cr
}}
$$

\noindent{\it Table 4. The $D=11$ supergravity cohomology, with respect to
the super-{\old3}-form.}

\section\Information{The meaning of the cohomologies}The maximally
supersymmetric theories we are considering have the property that
imposing the vanishing of certain lowest-dimensional field strengths,
($J$ and the $X$ tensors) implies the equations of motion.
This is related to the non-vanishing second cohomology.
The second cohomology contains a current supermultiplet, that enters
the field equations implied by the whole set of Bianchi identities
(that we do not consider in this paper). If the constraints for the
field strengths are changed, the equations of motion change.

To understand this, suppose for a moment that those cohomologies vanished.
Then, whatever constraint was put upon the field strengths (consistent
with the Bianchi identities) could be expressed as the image of
an exterior derivative acting on a {\old1}-form (a potential).
Such a field strength could be removed by a field redefinition, and
the system would be equivalent to one where those field strength components
were set to zero. One might ask whether it is not inconsistent to treat 
the field strength as belonging to the cohomology, since it is supposed
to be derived from a potential, and in that sense cohomologically trivial.
In fact, what the remaining Bianchi identities do is to resolve the
cohomology in the sense that the field strength indeed comes from a
potential. This is achieved by the equations of motion. Take \eg\ the
equation of motion for the spinor $\l$ in super-Yang--Mills. In the
undeformed case it reads $\Dslash\l=0$. When deformations are turned
on, it will read $\Dslash\l=\mu$, and if $\mu$ is derived from a
cohomologically trivial $J$, $\mu$ will be of the form $\mu=\Dslash\nu$.
Then the equation of motion for $\l$ is modified in a trivial way, 
removable by $\l-\nu\rightarrow\l$. Only if $J$ is cohomologically non-trivial
do the equations of motion receive significant modifications.
Once the equations of motion are taken into account, they state exactly
the integrability of the field strength to a potential (in the example
that $\mu$ is $\Dslash$ on something). So, the cohomology is resolved
by the field equations, but if it was trivial from the beginning, nothing
would have changed.

It is instructive to compare with a non-maximally supersymmetric gauge theory
known to possess an off-shell superfield formulation. Take the $\N=(1,0)$
super-Yang--Mills theory in $D=6$ [\NilssonSixDSYM].
This theory has an
off-shell formulation in terms of the vector, the spinor and a triplet
of auxiliary scalars of dimension 2. 
The complex is [\CederwallNilssonTsimpisI,\CederwallKarpacz]
$$
(000)(0)\Darrow0(100)(1)\Darrow1(200)(2)\Darrow2
\ldots\Darrow{n-1}(n00)(n)\Darrow{n}\ldots\Eqn\Complex
$$
Indeed, the only non-vanishing cohomologies are
$$
\eqalign{
\H^0&=(000)(0)_0\komma\cr
\H^1&=(010)(0)_1\oplus(001)(1)_{3\over2}\oplus(000)(2)_2\komma\cr
}\eqn
$$
where the representations are given as standard Dynkin labels for
Spin(1,5)$\times$SU(2) (the second factor being the $R$-symmetry group). 
The second cohomology is trivial, which also
is expected---setting $F_{\a\b}$
to zero does not put the theory on-shell, and the value of $F_{\a\b}$
does not contain any information about interactions---it can be set to
zero by a field redefinition.

It should be noted that even if this picture is quite clear for both
$D=10$ super-Yang--Mills and $D=11$ supergravity,
no explicit such understanding has been achieved for a formulation of
$D=11$ supergravity based on the super-{\old3}-form. We comment more
on this below.

There is a striking resemblance [\HowePrivate] between spinorial 
cohomology as it is constructed in this paper and the BRST cohomology
in Berkovits' covariant formulation of superstrings 
using pure spinors [\BerkovitsCohomology]. It seems as the choice of
representations building our complexes amounts to the same information
encoded in the contraction of the spinor derivative into a pure spinor.
Further investigation of this similarity, as well as the relation to
the pure spinors of ref. [\NilssonPure] should be pursued.

Reading the tables of cohomologies, a number of observations can be made.
Starting with the super-Yang--Mills case, table {\old1}, 
the interpretation is clear.
The gauge transformations (zeroth cohomology) and fields (first
cohomology) are the usual ones, and the second cohomology in the
field strength contains exactly the representations fitting into
the right hand sides of equations of motion for the spinor and vector.
This case has been worked out in full detail [\CederwallNilssonTsimpisI].
The only element of the cohomology that has not yet been explained
is the scalar of dimension {\old4} in the third cohomology.
It has the correct dimension for a lagrangian density, and we suspect 
that it might be related to an action principle containing the deformation
of the theory (not the ordinary kinetic terms).

In table {\old2}, the picture is similar. The first cohomology contains
the gauge transformations, the second one the five scalars, the
antisymmetric tensor and the two spinors, while the third cohomology
carries the representations of the currents: an (anti-)selfdual tensor
giving the selfduality of the {\old3}-form field strength, spinors of
the opposite chirality and five scalars, all at appropriate dimensions.

Turning to table {\old3}, and the $D=11$ supergravity,
the zeroth cohomology clearly represents the bosonic and fermionic
reparametrisations. In the first cohomology, we find the fields: at 
dimension 0 the (linearised) metric, at dimension $\half$ the gravitino
in (10001) and at dimension 1 the {\old4}-form field strength. In
addition, there is a spinor at dimension $\half$ and a vector at 
dimension 1. The experience from solving superspace Bianchi identities
for this system [\HoweWeyl,\CGNN,\GatesNishinoII]
tells us that these should be identified with the spinor and vector
components of the Weyl connection {\old1}-form\foot*{Although we use a
superspace that does not have Weyl scalings as part of its structure
group, we refer to these as Weyl connections, since they appear in
the torsion in exactly the places where they could be absorbed in a
Weyl connection by a conventional constraint.}. 
The second cohomology contains the representations for the ``usual'' equations 
of motion: $(10001)\oplus(00001)$ for the gravitino, $(00000)\oplus(20000)$
for the metric and (00100) for the {\old3}-form potential. Since the
latter one, due to gauge invariance, appears in the geometry only through
its field strength, one also finds its Bianchi identity in (00002) here.
In addition there are the representations $(01000)\oplus(10000)$ at
dimension 2.
One of the controversies over the possible deformations of the
eleven-dimensional supergravity theory concerns the r\^ole of the Weyl
connections. It has been conjectured that the corresponding curvatures
vanish, so that the they are integrable to a scalar field, the Weyl 
compensator, that can be removed by a conventional constraint
[\HoweWeyl,\CGNN], but also that they should play a significant r\^ole
in a deformed theory [\GatesNishinoII]. In ref. [\CGNN], we found
evidence for the first of these alternatives in the fact that the
spinor at dimension $\Fr32$ affects only the gravitino equation of motion,
not the Weyl curvature\foot\dagger{In that analysis, only the $\ss X$-tensor
in (10002), not the one in (11000), was included for simplicity.}. 
This of course does not follow from the present
listing of representations, but requires exact solution of the Bianchi
identities. It was also found that the dimension-1 part of the
Weyl curvature, $G_{\a\b}$, vanished, which looks natural in the absence
of any such cohomology. In the undeformed theory, all components of
the Weyl curvature vanish, but this has not been completely shown in a theory
with deformation by the $X$-tensors. We see that there {\it a priori} is
room in the second cohomology for a modification to the dimension-2
Weyl curvature $G_{ab}$ in (01000), but a treatment of the full set
of Bianchi identities [\CGNTprep] has to be awaited to determine whether it is
actually there.
From the present analysis, it is also not ruled out that there might
be corrections to the Bianchi identity for the {\old4}-form field strength.
Concerning the vector at dimension 2 we do not have any natural
interpretation, and find it plausible that it goes away when the full
set of Bianchi identities is considered.

Table {\old4}, containing the cohomologies for an antisymmetric tensor
in $D=11$ (the dimensions are adapted to a {\old3}-form potential), 
gives some new information. We are used to the fact that the tensor
field arises in a superspace formulation from super-geometry only, so
that the {\old4}-form field strength sits inside the torsion at
dimension 1, and that the closed super-{\old4}-form is constructed
out of geometrical data. 
The present analysis indicates that it is possible to turn this around,
and see the geometry as arising from the dynamics of a super-{\old3}-form
potential. 
It was observed in ref. [\CGNN] that the super-{\old4}-form, once the
deformations are turned on, is forced to have non-vanishing components
at negative dimensions. Here we will argue that these actually can
encode the deformations, and that this implies integrability conditions
on the $X$-tensors, not readily visible from the geometrical analysis
but presumably hidden therein.

The appearance of components at negative dimension immediately gives
rise to questions concerning propagation of branes and BPS conditions
in deformed backgrounds. Components of negative dimension in the 
brane Wess--Zumino terms have no counterpart in the kinetic term,
which complicates the issue of $\kappa$-symmetry [\Kappa]. Our opinion
is that the nature of $\kappa$-symmetry has to be changed in a deformed
theory, so that the local transformation parameter $\kappa$, seen as a
vector field on superspace, will have a non-vanishing Lorentz vector part.
In the embedding formalism [\HoweSezginWest], 
this would manifest itself as a deformation
of the embedding constraint.

In the second cohomology, representing gauge transformations with
parameter $\Lambda_{\a\b}$ in $(01000)\oplus(00002)$, 
we find the {\old2}-form gauge transformation
for the tensor field as well as bosonic and fermionic reparametrisations.
In the third cohomology, all the supergravity fields are contained in the
potential $C_{\a\b\g}$ in $(01001)\oplus(00003)$: at dimension
0 the {\old3}-form potential and the graviton, and at dimension $\half$
the gravitino. In addition, there is a scalar at dimension 0 and
a spinor at dimension $\half$. A tentative interpretation is that they
represent the Weyl compensator and the spinorial Weyl connection.
The fourth cohomology, sitting in the superfield $H_{\a\b\g\d}$ in 
$(02000)\oplus(01002)\oplus(00004)$, contains equations of motion for the
gravitino, for the metric and for the {\old3}-form, and no extra 
representations. Since now the {\old3}-form appears directly, and not via
its field strength, there is no need for its Bianchi identity in the
fourth cohomology. We observe that there is room neither for the vectorial 
Weyl connection not for its field strength. There is still a possibility
for an equation of motion for the spinor to mix in with the spinor part
of the gravitino equation of motion, but in the light of what the
geometrical analysis shows (see the previous paragraph) this seems very
unlikely.

It would be very interesting to analyse the $D=11$ supergravity 
from the point of view of the {\old3}-form instead of the super-geometry.
How this is done at a linearised level is obvious, but how  a full treatment
should be performed without invoking geometry is completely unclear,
although the present analysis indicates that it might be possible.
Such an approach might offer new perspectives on M-theory,
and we hope to be able to investigate it in the future.
One question that can be addressed in a traditional treatment is whether
or not the requirement that there exist a closed superspace {\old4}-form
puts stronger restrictions on the system than does a purely geometric
analysis. We are primarily aiming at the controversy about the fields
connected to Weyl scalings. Our guess is that the two approaches should
be equivalent, and that the Bianchi identities of higher dimensions
not considered in this paper enforce triviality of the Weyl bundle,
as shown in ref. [\HoweWeyl] for the undeformed theory and conjectured
in ref. [\CGNN] for the deformed theory. If this statement is true,
the {\old3}-form approach may actually offer some advantages over the
geometric picture, since it encodes the two-step 
integrability of the $X$-tensors to the three 
irreducible representations in $H_{\a\b\g\d}$ (this integrability goes
in another direction than the complex for the super-vielbein), something
that may be very useful when one wants to express these in terms of
physical fields in the supergravity multiplet to get explicit 
$\a'$-corrections from supersymmetry.
We believe that although the attempts made so far [\CGNN] have only been
partially successful, due to the technical complexity of the
calculations, a heavier use of computer techniques [\LiE,\GAMMA]
will render the investigations [\CGNTprep] tractable.

\pagegoal=30cm

\acknowledgements
MC wants to thank Itzhak Bars, Eric Bergshoeff,
Loriano Bonora and Kellogg Stelle for
discussions and comments. This work was supported in part by 
EU contract HPRN-CT-2000-00122 and by the Swedish Research Council.

\vfill\eject

\appendix{The LiE code}The code used with the program LiE to calculate
cohomologies in $D=10$ is:

\line{\hrulefill}

\noindent
{\tt
\null\hskip1.5cm \#\#\#\#\#\ definitions for D=10 SYM \#\#\#\#\#\hfill\break
\null\hskip1.5cm maxobjects 1000000				\hfill\break
\null\hskip1.5cm setdefault D5					\hfill\break
\null\hskip1.5cm rank=5						\hfill\break
\null\hskip1.5cm r(int n)=1X[0,0,0,n,0]				\hfill\break
\null\hskip1.5cm s=[0,0,0,1,0]					

\noindent
\null\hskip1.5cm \#\#\#\#\#\ calculate content of superfields \#\#\#\#\#
								\hfill\break
\null\hskip1.5cm r(int m,n)=					\hfill\break
\null\hskip1.5cm $\{$if n==0 then r(m) else			\hfill\break
\null\hskip1.5cm \null\hskip1cm	if n<0 then poly\_null(rank) else\hfill\break
\null\hskip1.5cm \null\hskip1cm	\null\hskip1cm	
		if m<0 then poly\_null(rank) else		\hfill\break
\null\hskip1.5cm \null\hskip1cm	\null\hskip1cm	\null\hskip1cm	
		tensor(r(m),alt\_tensor(n,s))			\hfill\break
\null\hskip1.5cm \null\hskip1cm	\null\hskip1cm	fi; 		\hfill\break
\null\hskip1.5cm \null\hskip1cm	fi; 				\hfill\break
\null\hskip1.5cm fi;						\hfill\break
\null\hskip1.5cm $\}$						

\noindent
\null\hskip1.5cm \#\#\#\#\#\ set negative multiplicities to zero \#\#\#\#\#
								\hfill\break
\null\hskip1.5cm pos\_pol(pol p)=				\hfill\break
\null\hskip1.5cm $\{$loc q=p;					\hfill\break
\null\hskip1.5cm for i=1 to length(p) do			\hfill\break
\null\hskip1.5cm \null\hskip1cm	if coef(p,i)<0 then q=q-p[i];	\hfill\break
\null\hskip1.5cm \null\hskip1cm	fi;				\hfill\break
\null\hskip1.5cm od;						\hfill\break
\null\hskip1.5cm q$\}$						

\noindent
\null\hskip1.5cm \#\#\#\#\#\ subtract multiplicities from the left \#\#\#\#\#
								\hfill\break
\null\hskip1.5cm left(int m,n)=					\hfill\break
\null\hskip1.5cm $\{$loc t=poly\_null(rank);			\hfill\break
\null\hskip1.5cm if m-1<0 then 					\hfill\break
\null\hskip1.5cm \null\hskip1cm	r(m,n);				\hfill\break
\null\hskip1.5cm else						\hfill\break
\null\hskip1.5cm \null\hskip1cm	t=r(0,m+n);			\hfill\break
\null\hskip1.5cm \null\hskip1cm	for k=1 to m do			\hfill\break
\null\hskip1.5cm \null\hskip1cm\null\hskip1cm		
				t=pos\_pol(r(k,m+n-k)-t);	\hfill\break
\null\hskip1.5cm \null\hskip1cm	od;				\hfill\break
\null\hskip1.5cm \null\hskip1cm	t;				\hfill\break
\null\hskip1.5cm fi						\hfill\break
\null\hskip1.5cm $\}$						

\noindent
\null\hskip1.5cm \#\#\#\#\#\ subtract multiplicities from the right \#\#\#\#\#
								\hfill\break
\null\hskip1.5cm right(int m,n)=				\hfill\break
\null\hskip1.5cm $\{$loc t=poly\_null(rank);			\hfill\break
\null\hskip1.5cm if n-1<0 then 					\hfill\break
\null\hskip1.5cm \null\hskip1cm	r(m,n);				\hfill\break
\null\hskip1.5cm else						\hfill\break
\null\hskip1.5cm \null\hskip1cm	t=r(m+n,0);			\hfill\break
\null\hskip1.5cm \null\hskip1cm	for k=1 to n do			\hfill\break
\null\hskip1.5cm \null\hskip1cm\null\hskip1cm		
				t=pos\_pol(r(m+n-k,k)-t);	\hfill\break
\null\hskip1.5cm \null\hskip1cm	od;				\hfill\break
\null\hskip1.5cm \null\hskip1cm	t;				\hfill\break
\null\hskip1.5cm fi						\hfill\break
\null\hskip1.5cm $\}$						

\noindent
\null\hskip1.5cm \#\#\#\#\#\ calculate candidate cohomologies \#\#\#\#\#
								\hfill\break
\null\hskip1.5cm h(int m,n)=pos\_pol(r(m,n)-right(m+1,n-1)-left(m-1,n+1))
}

\line{\hrulefill}

\noindent For the other three cases, the first part of the code (definitions)
is replaced by 

\line{\hrulefill}

\noindent 
{\tt
\null\hskip1.5cm \#\#\#\#\#\ definitions for D=6 tensor \#\#\#\#\#	\hfill\break
\null\hskip1.5cm maxobjects 1000000			\hfill\break
\null\hskip1.5cm setdefault A3C2				\hfill\break
\null\hskip1.5cm rank=5					\hfill\break
\null\hskip1.5cm s=1X[1,0,0,1,0]			

\noindent 
\null\hskip1.5cm \#\#\#\#\#\ build up reducible r(n) iteratively \#\#\#\#\#
							\hfill\break
\null\hskip1.5cm r(int n)=				\hfill\break
\null\hskip1.5cm $\{$loc q=poly\_null(rank);		\hfill\break
\null\hskip1.5cm if n\%2==0 then			\hfill\break
\null\hskip1.5cm \null\hskip1cm	loc k=n/2;		\hfill\break
\null\hskip1.5cm \null\hskip1cm	for i=0 to k do		\hfill\break
\null\hskip1.5cm \null\hskip1cm\null\hskip1cm		
				q=q+1X[n-2*i,i,0,n-2*i,i];	\hfill\break
\null\hskip1.5cm \null\hskip1cm	od;			\hfill\break
\null\hskip1.5cm fi;					\hfill\break
\null\hskip1.5cm if n\%2==1 then			\hfill\break
\null\hskip1.5cm \null\hskip1cm	loc k=(n-1)/2;		\hfill\break
\null\hskip1.5cm \null\hskip1cm	for i=0 to n/2 do	\hfill\break
\null\hskip1.5cm \null\hskip1cm\null\hskip1cm		
				q=q+1X[n-2*i,i,0,n-2*i,i];	\hfill\break
\null\hskip1.5cm \null\hskip1cm	od;			\hfill\break
\null\hskip1.5cm fi;					\hfill\break
\null\hskip1.5cm q;					\hfill\break
\null\hskip1.5cm $\}$					
 }

\line{\hrulefill}

\noindent by

\line{\hrulefill}

\noindent 
{\tt
\null\hskip1.5cm \#\#\#\#\#\ definitions for D=11 SG \#\#\#\#\#	\hfill\break
\null\hskip1.5cm maxobjects 1000000			\hfill\break
\null\hskip1.5cm setdefault B5				\hfill\break
\null\hskip1.5cm rank=5					\hfill\break
\null\hskip1.5cm s=1X[0,0,0,0,1]			

\noindent 
\null\hskip1.5cm \#\#\#\#\#\ build up reducible r(n) iteratively \#\#\#\#\#
							\hfill\break
\null\hskip1.5cm r(int n)=				\hfill\break
\null\hskip1.5cm $\{$loc q=poly\_null(rank);		\hfill\break
\null\hskip1.5cm if n\%2==0 then			\hfill\break
\null\hskip1.5cm \null\hskip1cm	loc k=n/2;		\hfill\break
\null\hskip1.5cm \null\hskip1cm	for i=0 to k do		\hfill\break
\null\hskip1.5cm \null\hskip1cm\null\hskip1cm		
				q=q+1X[1,i,0,0,n-2*i];	\hfill\break
\null\hskip1.5cm \null\hskip1cm	od;			\hfill\break
\null\hskip1.5cm fi;					\hfill\break
\null\hskip1.5cm if n\%2==1 then			\hfill\break
\null\hskip1.5cm \null\hskip1cm	loc k=(n-1)/2;		\hfill\break
\null\hskip1.5cm \null\hskip1cm	for i=0 to n/2 do	\hfill\break
\null\hskip1.5cm \null\hskip1cm\null\hskip1cm		
				q=q+1X[1,i,0,0,n-2*i];	\hfill\break
\null\hskip1.5cm \null\hskip1cm	od;			\hfill\break
\null\hskip1.5cm fi;					\hfill\break
\null\hskip1.5cm q;					\hfill\break
\null\hskip1.5cm $\}$					
 }

\line{\hrulefill}

\noindent and by

\line{\hrulefill}

{\tt
\noindent 
\null\hskip1.5cm \#\#\#\#\#\ definitions for D=11 tensor \#\#\#\#\#\hfill\break
\null\hskip1.5cm maxobjects 1000000			\hfill\break
\null\hskip1.5cm setdefault B5				\hfill\break
\null\hskip1.5cm rank=5					\hfill\break
\null\hskip1.5cm s=1X[0,0,0,0,1]			
					
\noindent 
\null\hskip1.5cm \#\#\#\#\#\ build up reducible r(n) iteratively \#\#\#\#\#
							\hfill\break
\null\hskip1.5cm r(int n)=				\hfill\break
\null\hskip1.5cm $\{$loc q=poly\_null(rank);		\hfill\break
\null\hskip1.5cm if n\%2==0 then			\hfill\break
\null\hskip1.5cm \null\hskip1cm	loc k=n/2;		\hfill\break
\null\hskip1.5cm \null\hskip1cm	for i=0 to k do		\hfill\break
\null\hskip1.5cm \null\hskip1cm\null\hskip1cm		
				q=q+1X[0,i,0,0,n-2*i];	\hfill\break
\null\hskip1.5cm \null\hskip1cm	od;			\hfill\break
\null\hskip1.5cm fi;					\hfill\break
\null\hskip1.5cm if n\%2==1 then			\hfill\break
\null\hskip1.5cm \null\hskip1cm	loc k=(n-1)/2;		\hfill\break
\null\hskip1.5cm \null\hskip1cm	for i=0 to n/2 do	\hfill\break
\null\hskip1.5cm \null\hskip1cm\null\hskip1cm		
				q=q+1X[0,i,0,0,n-2*i];	\hfill\break
\null\hskip1.5cm \null\hskip1cm	od;			\hfill\break
\null\hskip1.5cm fi;					\hfill\break
\null\hskip1.5cm q;					\hfill\break
\null\hskip1.5cm $\}$					
}

\line{\hrulefill}

\noindent respectively.
Since the code calculates possible cohomologies by subtraction
of multiplicities from left and right as described above, it will
produce a certain over-counting of mutually excluding possibilities.
For $D=10$, this type of ambiguity is not present.

\vfill\eject

\appendix{The content of $\N=1$, $D=10$ superfields}The ``vertical'' 
structure of an $\N=(1,0)$ scalar superfield in $D=10$ 
is as follows [\BergshoeffDeRoo]. 
Each row contains the irreducible content of the
completely antisymmetric product of $\ell$ chiral spinors
$\wedge^\ell(00010)$.

$$
\vtop{\baselineskip0pt\lineskip1pt
\ialign{
$\hfill#\quad$&$\,\hfill#\hfill\,$&$\,\hfill#\hfill\,$&$\,\hfill#\hfill\,$&$\,\hfill#\hfill\,$&$\,\hfill#\hfill$\cr
\ell=
 0&(00000)&    &       &    &       \cr
  &\sa    &    &       &    &       \cr
 1&(00010)&    &       &    &       \cr
  &\sa    &    &       &    &       \cr
 2&(00100)&    &       &    &       \cr
  &\sa    &    &       &    &       \cr
 3&(01001)&    &       &    &       \cr
  &\sa    &\sea&       &    &       \cr
 4&(10002)&    &(02000)&    &       \cr
  &\sa    &\sea&\sa    &    &       \cr
 5&(00003)&    &(11001)&    &       \cr
  &\sa    &\swa&\sa    &    &       \cr
 6&(01002)&    &(20100)&    &       \cr
  &\sa    &\swa&\sa    &    &       \cr
 7&(10101)&    &(30010)&    &       \cr
  &\sa    &\sea&\sa    &\sea&       \cr 
 8&(00200)&    &(20011)&    &(40000)\cr
  &\sa    &\swa&\sa    &\swa&       \cr 
 9&(10110)&    &(30001)&    &       \cr
  &\sa    &\sea&\sa    &    &       \cr
10&(01020)&    &(20100)&    &       \cr 
  &\sa    &\sea&\sa    &    &       \cr
11&(00030)&    &(11010)&    &       \cr
  &\sa    &\swa&\sa    &    &       \cr
12&(10020)&    &(02000)&    &       \cr
  &\sa    &\swa&       &    &       \cr
13&(01010)&    &       &    &       \cr 
  &\sa    &    &       &    &       \cr
14&(00100)&    &       &    &       \cr
  &\sa    &    &       &    &       \cr
15&(00001)&    &       &    &       \cr
  &\sa    &    &       &    &       \cr
16&(00000)&    &       &    &       \cr 
}}
$$

\vfill\eject

\refout

\end